\documentclass[a4paper]{article}
\usepackage[english]{babel}
\usepackage[utf8x]{inputenc}
\usepackage[T1]{fontenc}

\usepackage[a4paper,top=3cm,bottom=2cm,left=3cm,right=3cm,marginparwidth=1.75cm]{geometry}

\usepackage{amsmath}
\usepackage{graphicx}
\usepackage[colorinlistoftodos]{todonotes}
\usepackage[colorlinks=true, allcolors=blue]{hyperref}

\title{Mie scattering eigenmodes for optical trapping}
\author{Michael Mazilu\\
SUPA, School of Physics and Astronomy, \\
University of St Andrews, United Kingdom \\
E-mail: mm17@st-andrews.ac.uk}
\date{}

\begin{document}
\maketitle

\begin{abstract}
The Mie scattering theory enables the exact determination of the scattered field as a function of the incident field. Here, we use this approach to calculate the Hermitian  relationship between the incident field and the optical forces acting on the scattering objects. This Hermitian relationship defines also a set of orthogonal optical eigenmodes which deliver a natural basis to describe momentum transfer in light-matter interactions. 
\end{abstract}

\section{Introduction}

Scattering is one of the simplest light mater interactions possible. For spherical particles, this process can be described using the Lorenz-Mie theory, which makes use of vector spherical harmonic solutions of Maxwell’s equations to represent the fields involved. Using these solutions it is possible to describe the light field scattered from microscopic spherical particles and thus represent the field around a scattering object as a function of the incident fields. These solutions also allow us to determine the optical momentum transfer to the scattering object. This can be calculated using Maxwell stress tensor. In this paper, we will study the momentum transfer to a spherical particle illuminated by a superposition of vector Bessel beams and introduce the use of optical eigemodes to describe the optical forces acting of the microsphere.

\section{Theory}
In the following, we work with monochromatic fields ($\omega=k_0 c$) which define the incident field as a superposition of Bessel beams that are either azimuthally (s-polarisation)  or radially polarised (p-polarisation). Each Bessel beam is further characterised by its transversal wavevector $k_t=n_0 k_0 \sin(\gamma)$ and longitudinal wavevector $k_z=n_0 k_0 \cos(\gamma)$ where $\gamma$ and $n_0$ are the cone half-angle of the Bessel beam and index of refraction of the propagation media, respectively. 

In a first step, we define the vector Bessel beams in Cartesian coordinate system and their radial component in the spherical coordinate system. This component allows us to define the beam shape coefficients that together with the momentum transfer operator can be used to calculate the optical forces acting on spherical particles. The eigenmodes of the momentum operator can then be introduced, reducing the dimensionally of the system when calculating the momentum transfer. 

\subsection{Vector Bessel beam shape coefficients}

To simplify the definition of the Bessel beam in Cartesian coordinates, we introduce the following function:
\begin{eqnarray*}
B_\ell(k_t \rho)&=&(i)^\ell J_\ell(k_t \rho)e^{i\ell\phi}
\end{eqnarray*}
with $\rho=\sqrt{x^2+y^2}$ and $\phi=\arctan(y/x)$ and where $J_\ell(r)$ are the Bessel functions of the first kind.

In this case, the s-polarised Bessel beam takes the form:
\begin{equation*}
^{}{\bf E}^{\ell(s)}=\left(
\begin{array}{l}
^{}E^{\ell(s)}_x\\
^{}E^{\ell(s)}_y\\
^{}E^{\ell(s)}_z\\
\end{array}
\right)
=  c_w \left(
\begin{array}{l}
- i B_{\ell-1}(k_t \rho)+ i B_{\ell+1}(k_t \rho)\\
B_{\ell-1}(k_t \rho)+  B_{\ell+1}(k_t \rho)\\
0\\
\end{array}
\right)
\end{equation*}

\begin{equation*}
^{}{\bf H}^{\ell(s)}=\left(
\begin{array}{l}
^{}H^{\ell(s)}_x\\
^{}H^{\ell(s)}_y\\
^{}H^{\ell(s)}_z\\
\end{array}
\right)
=   c_w \frac{n}{c\mu_0}  \cos(\gamma) \left(
\begin{array}{l}
- B_{\ell-1}(k_t \rho)- B_{\ell+1}(k_t \rho)\\
- i B_{\ell-1}(k_t \rho)+i B_{\ell+1}(k_t \rho)\\
2  B_{\ell}(k_t \rho) \tan(\gamma) \\
\end{array}
\right)
\end{equation*}
where the $c_w=\pi \sin(\gamma)\sqrt{\cos(\gamma)} \exp(-i\omega t +i k_z z)$.

The p-polarised beam can be related to the s-polarised one via: $^{}{\bf E}^{\ell(p)}=-c\mu_0/n ^{}{\bf H}^{\ell(s)}$ and $^{}{\bf H}^{\ell(p)}=n/(c\mu_0) ^{}{\bf E}^{\ell(s)}$. The main property of these Bessel beams is that they are eigenfunctions of the solid rotation operator with a integer valued eigenvalue $\ell$ 
\begin{eqnarray*}
-i{\bf e}_z\times{\bf E}^{\ell(s,p)}-i\left( {\bf r }\times\nabla\right)_{{\bf e}_z}{\bf E}^{\ell(s,p)}&=&\ell{\bf E}^{\ell(s,p)}\\
-i{\bf e}_z\times{\bf H}^{\ell(s,p)}-i\left( {\bf r }\times\nabla\right)_{{\bf e}_z}{\bf H}^{\ell(s,p)}&=&\ell{\bf H}^{\ell(s,p)}
\end{eqnarray*}
where ${\bf e}_z$ is the z-direction unit vector. The two terms in this operator can be identified with the spin angular momentum and the orbital angular momentum, respectively. 

Projecting the electric and magnetic fields on the radial unit vector in spherical coordinates defines the radial components of the electromagnetic fields as
\begin{eqnarray*}
^{}E^{\ell(s)}_r(r,\theta,\phi)&=&  i^\ell c_s  \exp(i\ell \phi)\left(
-J_{\ell-1}(k_t r \sin\theta)\sin\theta- J_{\ell+1}(k_t r \sin\theta) \sin\theta
\right) \\
^{}H^{\ell(s)}_r(r,\theta,\phi)&=&  i^\ell  c_s  \frac{n}{c\mu_0}  \exp(i\ell \phi)\\
&&(
 2J_{\ell}(k_t r \sin\theta) \cos\theta\sin\gamma+i J_{\ell-1}(k_t r \sin\theta) \sin\theta\cos\gamma  -iJ_{\ell+1}(k_t r \sin\theta) \sin\theta\cos\gamma)
\end{eqnarray*}
where
$c_s=\pi \sin(\gamma)\sqrt{\cos(\gamma)} \exp(-i\omega t +i k_z z \cos(\theta))$

The beam shape coefficients can then be determined by calculating the inner product of the radial field component with the spherical harmonic function $Y^m_n(\theta,\phi)$

\begin{equation*}
j_n(kr)\left(
\begin{array}{l}
-g^{TM}_{nm}\\
g^{TE}_{nm}/Z
\end{array}
\right)=\frac{kr}{\sqrt{n(n+1)}}
\int_0^\pi d\theta\int_0^{2 \pi}d\phi \;\;{Y^m_n}(\theta,\phi)^*
\left(
\begin{array}{l}
^{}E_r\\
^{}H_r
\end{array}
\right)
\end{equation*}
with $Z=\mu_0 c/n_0$ and where $*$ stands for the complex conjugate and where $g^{TM}_{nm}$ and $g^{TE}_{nm}$ are the beam shape coefficients. 

Introducing
$c_m=\pi \sin(\gamma)\sqrt{\cos(\gamma)} \exp(-i\omega t )$, we have for the s-polarised Bessel beam
\begin{eqnarray*}
g^{TM}_{nm}&=& \delta_m^\ell \frac{ 4 \pi i^{n} \sqrt{\pi\cos\gamma(2n+1)(n-m)!}}{\sqrt{ n(n+1)(n+m)!}} (m)P^m_n(\cos\gamma)\\
g^{TE}_{nm}&=&\delta_m^\ell
\frac{4 \pi  i^n \sqrt{\pi \cos\gamma(2n+1)(n-m)!}}{\sqrt{ n(n+1)(n+m)!}} (-i)
((n+1)\cos\gamma P^m_n(\cos\gamma)+(m-n-1)P^m_{n+1}(\cos\gamma))\\
\end{eqnarray*}
valid for $n\ge \ell$.
In the following, we define a single list of beam shape coefficients $$g_j=g^{p_j}_{n_jm_j}$$
where $m_j$, $n_j$ and $p_j$ are all coefficients considered indexed using the subscript $j$. The polarisation index $p_k$ can take the values of 1 and 2 corresponding respectively to $TM$ and $TE$ modes. A superposition of vector Bessel beams will then correspond to a superposition of beam shape coefficients $g_i$. 
Indeed, an angular dependent  superposition of Bessel beams defined this way makes it  possible to describe high-NA beams used in microscopes which introduce  spherical aberrations~\cite{Neves:2007p6055}.

\subsection{Momentum transfer}
Momentum transfer to scattering objects can be calculated using Maxwell's stress tensor defined by
\begin{eqnarray*}
\widetilde{\sigma} 
&=&\frac{1}{2} \left(\left(n_0^2\epsilon_{0}{\bf E}\cdot\mathbf{E}+\mu_{0}{\bf H}\cdot{\bf H}\right)\widetilde{I}-2 n_0^2\epsilon_{0}{\bf E}\otimes{\bf  E} -2\mu_{0}\mathbf{H}\otimes\mathbf{H}\right),
\end{eqnarray*}
 where $\otimes$ stands for the tensor product and $\widetilde{I}$
for the identity 3x3 matrix. 
  In a first instance, we are interested in the force in the $z$-direction which can be calculated by integrating over a spherical surface $S$ surrounding the object
\begin{equation}\label{integFz}
F_z=<\oint_S \mathbf{e}_z\widetilde{\sigma} d\mathbf{n} >
\end{equation} 
where $<>$ stands for the time average over the optical cycle. The fields $\mathbf{E}$ and $\mathbf{H}$ correspond to the sum of the incident and scattered fields. The incident fields are defined by the beam shape coefficients while the scattered fields can be calculated using the Mie scattering coefficients and the beam shape coefficients. Altogether, we remark that the optical force acting in the z-direction can be expressed in a quadratic form with respect to the beam shape coefficients.  This quadratic Hermitian form is based on the matrix operator
\begin{eqnarray*}
M_k^j=\frac{1}{2}
\delta_{m_k}^{m_j}\delta_{p_k}^{p_j} 
\left(\delta_{n_k}^{n_j+1}\sqrt{\frac{(1-n_k)^2(n_j^2-m_k^2)}{n_k^2(4n_k^2-1)}}-\delta_{n_k+1}^{n_j}\sqrt{\frac{(1-n_j)^2(n_k^2-m_k^2)}{n_j^2(4n_j^2-1)}}\right)\\
\left(\delta_{p_k}^1(b_k+b^*_j-2b_kb_j^*) +\delta_{p_k}^2 (a_k+a^*_j-2a_ka_j^*) \right) \\
+\frac{i}{2}\delta_{n_k}^{n_j} \delta_{m_k}^{m_j}
\frac{m_k}{n_k(n_k+1)}\left(\delta_{p_k}^{1} \delta_{2}^{p_j} (b_k+a_k^*-2b_ka_k^*)-\delta_{p_k}^{2} \delta_{1}^{p_j} (a_k+b_k^*-2a_kb_k^*) \right)
\end{eqnarray*}
where $n_k$ $m_k$ and $p_k$ is the $n$- and $m$-index and polarisation state of the $k$-th beam shape component and where $a_k$ and $b_k$ are the $n_k$-indexed Mie scattering coefficients of the scattering object.  

The optical momentum transfer in the z-direction is then defined by 
\begin{equation}\label{QF}
F_z=g^kM_k^jg_j
\end{equation}
where $g^k=g_k^*$. This quadratic Hermitian expression calculates the force acting in the z-direction on the scattering object positioned in at the origin of the coordinate system however the direction can be changed using spherical harmonics rotation and translation matrices.

\subsection{Optical eigenmodes}
The momentum transfer matrix is by construction Hermitian and in general all real field properties that can be expressed as a function of field in a quadratic form will lead to a matrix/operator that is Hermitian~\cite{Baumgartl:2011bm,DeLuca:2011jl}. The consequence of this observation is that the momentum transfer matrix defines a set of orthogonal vectors that correspond to the eigenvectors of $M_k^j$. Each of these eigenvectors defines a field that we call optical eigenmode of the measure in questions, in this case it is the optical eigenmode of $F_z$. If the Mie scattering order tends to infinity then the optical eigenmodes will form a complete Hilbert basis set of solutions of Maxwell's equations. 
Each quadratic  measure will define an operator with its own set of optical eigenmodes. If two operators commute then it is possible to define a set of optical eigenmodes that are simultaneously eigenmodes for each operator.  

\begin{figure}[htbp] 
   \centering
   \includegraphics[width=10cm]{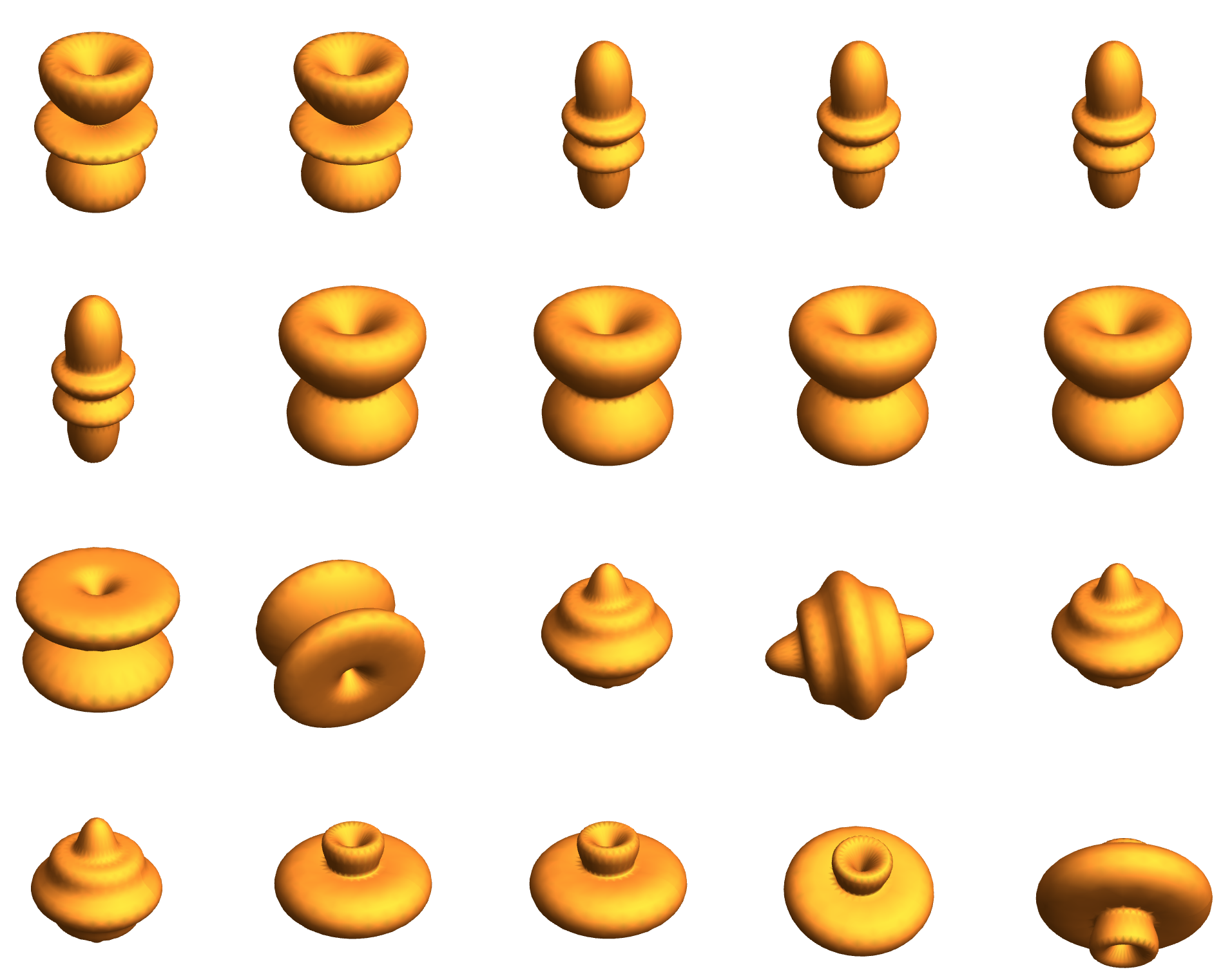}
   \caption{Angular intensity distribution of the first 20 optical eigenmodes.}
   \label{fig1}
\end{figure}

Each optical eigenmodes has a real eigenvalue associated with it. This value corresponds in this case to the optical force acting in the z-direction on the scattering object when the incident field is the optical eigenmode. Considering the set of all operators that commute with each other then the set of  eigenvalues from these operators can be used to uniquely identify an optical field. 

Further, the eigenvalues can be used to sort the optical eigenmodes by order of importance  such that we can reduce the degrees of freedom for which a system needs to be solved for by discarding all optical eigenmodes that are not contributing. It is this property that can be used to improve the computational speed of the numerical model. We determine the optical eigenmodes that lead to any significant momentum transfer and use these modes to describe the incident field. Any incident field that does not couple to the optical eigenmodes will not lead to any measurable momentum transfer due to the symmetries of the problem considered. 

\section{Applications}

The procedure outlined above can be used to calculate the optical trapping strength of trapped homogeneous or coated micro-particles including optical aberration introduced by high-NA microscope objectives\cite{Craig:2015kn}. 
Optical forces for larger particles can also be calculated and it is possible to determine the complex trajectories of levitated micro-particles in vacuum\cite{Mazilu:2016jn}.


In the following, we highlight the linear momentum optical eigenmodes in the simple case of a spherical particle being one wavelength in diameter with an index of refraction of 1.5 in vacuum. Figure~\ref{fig1} shows the angular distribution of the intensity of the z-direction optical eigenmodes. The figures are order by the magnitude of the linear momentum transferred.  


\end{document}